\RequirePackage{fix-cm}
\documentclass[twocolumn,epjc3]{svjour3}  
\usepackage{amsmath}
\smartqed  
\RequirePackage{graphicx}
 \RequirePackage{mathptmx}      
\RequirePackage{latexsym}
\RequirePackage[numbers,sort&compress]{natbib}
\journalname{Eur. Phys. J. C}
\begin{document}

\title{Metric-affine bumblebee gravity: classical aspects
}


\author{Adri\`a Delhom\thanksref{e1,addr1}
        \and
        J. R. Nascimento\thanksref{e2,addr2} 
        \and
        Gonzalo J. Olmo\thanksref{e3,addr1,addr2}
        \and
        A. Yu. Petrov\thanksref{e4,addr2}
        \and
        Paulo J. Porf\'{i}rio\thanksref{e5,addr2}
}

\thankstext{e1}{e-mail: adria.delhom@uv.es}
\thankstext{e2}{e-mail: jroberto@fisica.ufpb.br}
\thankstext{e3}{e-mail: gonzalo.olmo@uv.es}
\thankstext{e4}{e-mail: petrov@fisica.ufpb.br}
\thankstext{e5}{e-mail: pporfirio@fisica.ufpb.br}


\institute{Departament de F\'{i}sica Te\`{o}rica and IFIC, Centro Mixto Universitat de
Val\`{e}ncia - CSIC, 
Universitat de Val\`{e}ncia, Burjassot-46100, Val\`{e}ncia, Spain\label{addr1}
           \and
           Departamento de F\'{\i}sica, Universidade Federal da 
Para\'{\i}ba, 
 Caixa Postal 5008, 58051-970, Jo\~ao Pessoa, Para\'{\i}ba, Brazil\label{addr2}
}

\date{Received: date / Accepted: date}

\maketitle

\begin{abstract}
We consider the metric-affine formulation of bumblebee gravity, derive the field equations, and  show that the connection can be written as Levi-Civita of a disformally related metric in which the bumblebee field determines the disformal part. As a consequence, the bumblebee field gets coupled to all the other matter fields present in the theory, potentially leading to nontrivial phenomenological effects. To explore this issue we compute the post-Minkowskian, weak-field limit  and study the resulting effective theory. In this scenario, we couple scalar and spinorial matter to the effective metric, and then we explore the physical properties of the VEV of the bumblebee field, focusing mainly on the dispersion relations and the stability of the resulting effective theory.
\keywords{Symmetry breaking \and bumblebee model \and metric-affine gravity}
\end{abstract}

\section{Introduction}
\label{sec:intro}

The consistent inclusion of Lorentz symmetry breaking in a curved space is certainly one of the most important open problems within studies of this phenomenological idea.  Traditionally, the Lorentz symmetry breaking has been introduced in two manners, 1) the explicit one, where a constant vector (tensor) is introduced in the theory from the very beginning, and 2) the spontaneous one, where this constant vector (tensor) arises as the vacuum expectation value of some dynamical field (see \cite{KosGra} for a detailed discussion of approaches to Lorentz symmetry breaking in gravity).  

It should be { emphasized} that while the explicit approach is very convenient in the case of linearized gravity, there are essential difficulties with its application in a full-fledged scenario. Indeed, unlike in flat space, in curved space it is highly problematic to define constant vectors (or tensors) because the condition of vanishing covariant derivatives of a given tensor imposes constraints on the background metric which are difficult or even impossible to satisfy. Otherwise, quantum corrections in curved-space extensions of known Lorentz-breaking field theory models will involve an infinite tower of new terms proportional to covariant derivatives of ``constant" Lorentz-breaking tensors, which makes the calculations much more complicated (see {\it e.g.} \cite{ShaBP}). Another difficulty related with this approach is that in a curved space-time, the group of general covariant transformations plays a double role, being not only the extension of the Lorentz group but also the gauge group. As a result, introducing terms that explicitly break Lorentz symmetry will imply violation of the gauge symmetry as well, which makes the spontaneous symmetry breaking approach much more appropriate (that see discussion in \cite{KosMew}).

 In {a curved space-time}, the most convenient form of { introducing} spontaneous Lorentz symmetry breaking is the bumblebee model, first  proposed  in \cite{KosGra}, where the breaking mechanism is implemented via a new dynamical vector field with a nontrivial potential characterized by a {continuous set of minima}. The first studies of modifications of known gravitational solutions within the bumblebee gravity have been carried out in \cite{Bertolami}. 
Various aspects of the bumblebee gravity have been studied in numerous papers, {\it e.g.} \cite{Seifert,Maluf,Godel,Casana:2017jkc}, all of which assume that the underlying geometry is of (pseudo)-Riemannian type. However, in the absence of any empirical evidence supporting that the space-time structure is { necessarily} Riemannian, or otherwise\footnote{Here Riemannian means that the connection is completely specified by the metric, as opposed to being an {\it a priori } independent geometrical entity.}, in the high energy regime  \cite{Will:2014kxa,Berti:2015itd,Daniel:2010ky,Krawczynski:2012ac,Damour:1991rd,Kramer:2006nb,Clifton:2005aj,Abbott:2016blz}, it is legitimate to explore other alternatives. In particular, it is well-known that the metric-affine (or Palatini) formulation of gravity theories beyond GR yields field equations which are inequivalent to those obtained in the purely metric approach, and it has been argued that quantum effects could be encoded at low energies by a non-Riemannian counterpart of the space-time geometry \cite{Olmo:2015bha,Olmo:2015wwa,Lobo:2014nwa,Hossenfelder:2017rub}. This fact motivates us to initiate the exploration of Lorentz breaking phenomenology of bumblebee gravity in its metric-affine formulation, in which the connection is treated as a geometrical object a priori independent of the metric. In this sense, it is worth mentioning that Lorentz-breaking extensions of both the Standard Model and General Relativity have already been tested experimentally, leading to stringent constraints on Lorentz-breaking parameters, see {\it e.g.} \cite{datatable,Kostelecky:2015dpa,Bailey:2014bta,Kostelecky:2016kfm,Kostelecky:2010ze,Kostelecky:2008in,Bailey:2006fd}.

It has been recently shown that some metric-affine theories of gravity beyond GR have a peculiar behavior that makes them depart from their metric counterpart and become potentially testable via elementary particle interactions. In the particular case of minimally coupled Ricci-Based Gravity theories (RBG's), those in which the gravity Lagrangian is a (projective invariant) function of the metric and the Ricci tensor, one finds that the space-time metric picks up two types of contributions, one coming from the integration over the matter sources and which is mainly responsible for the space-time curvature, and another coming from the energy-momentum density of the local sources. The latter contribution is responsible for the existence of a nonzero non-metricity tensor, $Q_{\mu\alpha\beta} =\nabla_\mu g_{\alpha\beta}$, in regions where the stress-energy of the matter fields is not covariantly constant. The origin of these contributions in the metric can be traced back to the fact that these theories admit an Einstein frame (via a non-conformal transformation) in which the matter sector features new non-linear interactions and the gravity sector is described by GR. Thus, from the original frame of the theory (RBG frame), these non-linear interactions appear encoded in the space-time metric and are responsible for the non-vanishing non-metricity tensor. The existence of the Einstein frame representation for RBG theories was devised in \cite{Afonso:2017bxr,BeltranJimenez:2017doy}, and it has been explicitly proven and used for applications for different matter sectors: perfect and anisotropic fluids, scalar fields and non-linear electrodynamics \cite{Afonso:2018bpv,Afonso:2018hyj,Afonso:2018mxn,Afonso:2019fzv,Delhom:2019zrb}. 
The deviation from the Riemannian condition $\nabla_\mu g_{\alpha\beta}=0$ in the RBG frame is intimately related to a departure of the space-time metric from its Minkowskian form whenever matter fields are present even if the effects of curvature ({\it i.e.} Newtonian and post-Newtonian corrections) are negligible. As a result, particle physics experiments can be used to place strong constraints on the parameters of those gravity models \cite{Latorre:2017uve,Delhom:2019wir}. 

In this work we will consider a so far unexplored route, in which the bumblebee model plays a key role. In this model the gravity Lagrangian is represented by the Einstein-Palatini GR Lagrangian and an additional (projectively invariant) non-minimal coupling term between the bumblebee field and the affine Ricci tensor. This sort of non-minimal coupling has not yet been considered in detail within metric-affine theories. In addition, keeping the gravitational part exactly as in GR avoids the generation of non-metricity induced by the stress-energy tensor of the matter fields, which allows to disentangle the different contributions that the non-metricity may have. As we will see, the resulting theory admits an exact formal solution for the independent connection which leads to the emergence of a non-metricity tensor generated by the non-minimal bumblebee coupling. The bumblebee field will then become coupled to the rest of the matter fields present in the theory due to its non-minimal coupling with the gravitational sector. Thus, even though the theory is initially formulated following the postulates of metric theories of gravity so as to satisfy the Einstein equivalence principle \cite{Will}, in the end the resulting coupling of the bumblebee field to the matter sector implies a direct violation of this principle, as one would expect in a Lorentz violating theory. Besides deriving the field equations and discussing their resolution, in this paper we will also discuss the weak field, post-Minkowskian limit with a focus on the modified dispersion relations and stability of scalar and Dirac fields. 

The structure of our paper is as follows. In section 2, we present a general family of Ricci-Based Gravity theories (RBGs) with non-minimal couplings between matter and the connection, introducing the  metric-affine bumblebee model as a specific implementation. After discussing the field equations of the bumblebee model, in section 3, we study the weak-field, post-Minkowskian limit and the effective dynamics when curvature effects are negligible as compared to the contribution of non-metricity. Section 4 is devoted to the derivation and discussion of the modified dispersion relations and stability corresponding to scalar and spinor fields. We conclude with a summary and discussion of the results.

\section{Metric-affine Bumblebee model as a non-minimally coupled RBG}

\subsection{General case in the RBG framework.}

The analysis of metric-affine RBG theories has focused  so far on modifications of the gravitational sector but keeping a minimally coupled matter sector, {\it i.e.} with the matter fields interacting only with the metric, not with the connection. Such theories can be described by an action of the form
\begin{eqnarray} 
\mathcal{S}=\frac{1}{2\kappa^2}\int d^4 x\,\sqrt{-g}F(g^{\mu\nu},R_{(\mu\nu)}(\Gamma))+ \mathcal{S}_{m}(g_{\mu\nu},\Psi_i) \ ,
\label{MinCoupRBG0}
\end{eqnarray}
where $\kappa^2=8\pi G$, $\Psi_i$ denotes a collection of minimally coupled matter fields, and $R_{(\mu\nu)}(\Gamma))$ represents the symmetrized Ricci tensor of the independent connection $\Gamma^\alpha_{\mu\nu}$. We use units $\hbar=c=1$.  A straightforward way of {generalizing} the above action is to allow for a non-minimal coupling between some matter fields $\Phi_i$ and the connection via $R_{(\mu\nu)}(\Gamma))$, such that (\ref{MinCoupRBG0}) turns into
\begin{eqnarray} 
\nonumber\mathcal{S}&=&\frac{1}{2\kappa^2}\int d^4 x\,\sqrt{-g}F(g^{\mu\nu},R_{(\mu\nu)}(\Gamma),\Phi_i)+ \mathcal{S}_{m}(g_{\mu\nu},\Psi_i),\\
\label{MinCoupRBG}
\end{eqnarray}
where the $\Phi_i$ have minimal kinetic terms in the sense of \cite{Delhom:2020hkb}. The detail that only the symmetrized part of the Ricci tensor enters the action guarantees the existence of a projective symmetry, which turns out to be relevant for the stability of the theory by ensuring the absence of ghost-like degrees of freedom in RBG theories \cite{BeltranJimenez:2019acz,Jimenez:2020dpn} as well as in metric-affine scalar-tensor theories \cite{Aoki:2019rvi}.

In the metric-affine framework, metric and connection are treated as independent fundamental fields. Accordingly,  their field equations are obtained by extremizing the action without imposing any {\it a priori} relation between $\delta g_{\mu\nu}$ and $\delta \Gamma^\alpha_{\mu\nu}$. By varying upon the above action, one finds that the metric and connection field equations of the RBG action with non-minimally coupled matter fields are formally identical to those of the minimally coupled version, namely
\begin{eqnarray}
\frac{\partial F}{\partial g^{\mu\nu}}-\frac{1}{2}F g_{\mu\nu}&=&\kappa^2 T^M_{\mu\nu};\\
\nonumber\nabla^\Gamma_\lambda \left(\sqrt{-h}h^{\mu\nu}\right)&=&\sqrt{-h}\bigg[T^{\mu}{}_{\lambda\alpha}h^{\nu\alpha}+T^{\alpha}{}_{\alpha\lambda}h^{\mu\nu}+\\
&+&\frac{1}{3}T^{\alpha}{}_{\alpha\beta}\delta^\mu{}_{\lambda}\bigg]\label{conneqRBG},
\end{eqnarray}
where $T^{\alpha}{}_{\mu\nu}=2\Gamma^{\alpha}{}_{[\mu\nu]}$  is the torsion tensor, $T^M_{\mu\nu}$ is the usual stress-energy tensor of the minimally-coupled matter sector $T_{\mu\nu}^{M}=-\frac{2}{\sqrt{-g}}\frac{\delta(\sqrt{-g}\mathcal{L}_{M})}{\delta g^{\mu\nu}}$, and we have defined a new (inverse) metric $h^{\mu\nu}$ via $\sqrt{-h}h^{\mu\nu}=2\kappa^2 \sqrt{-g}\frac{\partial F}{\partial R_{(\mu\nu)}}$. Following \cite{BeltranJimenez:2017doy}, it is possible to show that the torsion terms on the right-hand side of (\ref{conneqRBG}) can be eliminated by exploiting the projective symmetry of the theory. After doing this, one finds that, up to a {non-physical} projective mode, the solution for the connection is given by the Levi-Civita connection of the metric $h_{\mu\nu}$, namely
\begin{equation}
{\Gamma}^{\mu}{}_{\alpha \beta}=\frac{1}{2}h^{\mu \lambda}\left(\partial_{\alpha} h_{\beta \lambda}+\partial_{\beta} h_{\lambda \alpha}-\partial_{\lambda} h_{\alpha \beta}\right) \ .
\end{equation}
We then see that the inclusion of matter fields that are non-minimally coupled through the Ricci tensor is still compatible with the fact that in RBG theories the connection is the Levi-Civita connection of some metric, which greatly simplifies the process of solving the field equations.

\subsection{Building the Einstein frame of generalized RBGs with matter couplings to $R_{\mu\nu}$.}

The fact that the connection can be solved as the Levi-Civitta connection of a metric $h_{\mu\nu}$ raises the question of whether these generalised RBGs also admit an Einstein frame representation. As we show next, the answer is positive and the Einstein frame can be constructed with the standard procedure (see {\it e.g.} \cite{BeltranJimenez:2017doy}) of linearising the action by introducing an auxiliary (symmetric) tensor field $\Sigma_{\mu\nu}$. 
 Let us sketch how the Einstein-frame is reached for a generalised RBG of the form (\ref{MinCoupRBG}). Consider the action 
\begin{eqnarray} 
\nonumber\tilde{\mathcal{S}}&=&\frac{1}{2\kappa^2}\int d^4 x\,\sqrt{-g}\left[\frac{\delta f}{\delta \Sigma_{\mu\nu}}(R_{(\mu\nu)}-\Sigma_{\mu\nu})+f \right]+\\
&+& \mathcal{S}_{m}(g_{\mu\nu},\Psi_i) \ ,
\label{MinCoupRBGauxiliary}
\end{eqnarray}
where $f=F(g^{\mu\nu},\Sigma_{\mu\nu},\Phi_i)$ indicates that we have replaced all instances of $R_{(\mu\nu)}$ by $\Sigma_{\mu\nu}$. The field equations of $\Sigma_{\mu\nu}$ are algebraic, and imply that $\Sigma_{\mu\nu}=R_{(\mu\nu)}$ provided that $\frac{\delta^2 f}{\delta\Sigma^2}\neq0$. Hence, integrating out $\Sigma_{\mu\nu}$ we see that the {\it new} action (\ref{MinCoupRBGauxiliary}) is physically equivalent to the original one (\ref{MinCoupRBG}). 

Let us now define the inverse metric $h^{\mu\nu}$ by 
\begin{equation}
\sqrt{-h}h^{\mu\nu}\equiv \sqrt{-g} \frac{\delta f}{\delta \Sigma_{\mu\nu}}.
\end{equation}
Generally, we can see the above equation as an implicit definition of $\Sigma_{\mu\nu}$, which should be seen as a function of $(g_{\mu\nu},h_{\mu\nu},$ $\Phi_i)$. Thus, by solving algebraically in $\Sigma_{\mu\nu}$, we would be able to write $\Sigma_{\mu\nu}(g^{\mu\nu},h^{\mu\nu},\Phi_i)$. We can then perform a field redefinition in (\ref{MinCoupRBGauxiliary}) 
 by substituting $\Sigma_{\mu\nu}(g^{\mu\nu},h^{\mu\nu},\Phi_i)$ to arrive at
 \begin{eqnarray} 
\nonumber\tilde{\mathcal{S}}&=&\frac{1}{2\kappa^2}\int d^4 x\,\left[\sqrt{-h}h^{\mu\nu}R_{\mu\nu}+{\mathcal{U}}(g^{\mu\nu},h^{\mu\nu},\Phi_i) \right]+\\
 &+&\mathcal{S}_{m}(g_{\mu\nu},\Psi_i),
\label{MinCoupRBGalmostEF}
\end{eqnarray}
where
\begin{eqnarray} 
\nonumber &&{\mathcal{U}}(g^{\mu\nu},h^{\mu\nu},\Phi_i)\equiv
\\&=&
\sqrt{-g}\left(f\left.-\frac{\delta f}{\delta\Sigma_{\mu\nu}}\Sigma_{\mu\nu}\right)\right|_{\Sigma_{\mu\nu}=\Sigma_{\mu\nu}(g^{\mu\nu},h^{\mu\nu},\Phi_i)}.\nonumber
\end{eqnarray}
Now, we see that the field equations for $g^{\mu\nu}$ are also algebraic and, therefore, we can formally solve them to obtain $g^{\mu\nu}(h^{\mu\nu},\Phi_i,\Psi_i)$. Thus, integrating $g^{\mu\nu}$ out by means of this solution we arrive to the Einstein-Hilbert representation of (\ref{MinCoupRBG}), given by
\begin{equation}
{\mathcal{S}}_{\rm EH}=\frac{1}{2\kappa^2}\int d^4 x\,\sqrt{-h}h^{\mu\nu}R_{\mu\nu}+ \bar{\mathcal{S}}_{m}(h_{\mu\nu},\Phi_i,\Psi_i),\label{EFaction}
\end{equation}
where 
\begin{eqnarray}
\nonumber\bar{\mathcal{S}}_{m}(h_{\mu\nu},\Phi_i,\Psi_i)&=&{\mathcal{S}}_{m}\big[g^{\mu\nu}(h^{\mu\nu},\Phi_i,\Psi_i),\Psi_i\big] +\\
\nonumber &+& \int d^4 x\,{\mathcal{U}}\big[g^{\mu\nu}(h^{\mu\nu},\Phi_i,\Psi_i),h^{\mu\nu},\Phi_i\big].
\end{eqnarray}
 We can clearly see here that the above action (\ref{EFaction}), which by construction is equivalent to \ref{MinCoupRBG}, formally describes a set of Einstein-like equations for the metric $h_{\mu\nu}$ coupled to a matter sector described by $\bar{\mathcal{S}}_{m}$ which will feature {\it new} interactions between the $\Phi_i$ and $\Psi_i$ sectors.

\subsection{The metric-affine bumblebee model as a particular case.}

As a particular case of the class of theories discussed above, we find the metric-affine version of the curved space-time bumblebee model (see \cite{Casana:2017jkc}), which is defined by an action of the form 
\begin{eqnarray} 
\nonumber\mathcal{S}&=&\int d^4 x\,\sqrt{-g}\Big[\frac{1}{2\kappa^2}\Big(R(\Gamma)+\xi B^{\alpha} B^{\beta} R_{\alpha\beta}(\Gamma)\Big)-\\
&-&\frac{1}{4}B^{\mu\nu}B_{\mu\nu}-V(B^{\mu}B_{\mu}\pm b^2)\Big] + \mathcal{S}_{m}(g_{\mu\nu},\Psi_i) \ .
\label{bumblebee}
\end{eqnarray}
 A key feature of this model is that the bumblebee field $B_\mu$ has a non-zero  vacuum expectation value (VEV) that spontaneously breaks Lorentz symmetry by introducing a privileged space-time direction. This field is coupled non-mini- mally to the space-time geometry via { the} $B^\mu B^\nu R_{\mu\nu}$ term. The parameter $\xi$ characterizes the strength of this non-mini- mal coupling between the bumblebee field and the affine connection through $R_{\mu\nu}(\Gamma)$, and we use the minimal coupling prescription for its kinetic term, so that the field strength of the bumblebee field is  defined as the exterior derivative of $B_{\mu}$ ({\it i.e.} $B_{\mu\nu}=(\mbox{d}B)_{\mu\nu}$). 

We see that by identifying $\Phi_i\equiv B_\mu$ we find
\begin{align}
& f_B=(g^{\mu\nu}+\xi B^\mu B^\nu)\Sigma_{\mu\nu},\\
& \sqrt{-h}h^{\mu\nu}=g^{\mu\nu}+\xi B^\mu B^\nu,\label{auxiliarymetric0}\\
& \mathcal{U}_B=-\frac{1}{4}B^{\mu\nu}B_{\mu\nu}-V(B^\mu B_\mu\pm b^2),\label{auxiliarypotential}
\end{align}
which would allow us to write the Einstein-frame action corresponding to \ref{bumblebee} once the metric $g_{\mu\nu}$ is solved in terms of the matter fields $(B_\mu, \Psi_i)$.

 By taking variations of Eq. (\ref{bumblebee}) with respect to the metric, the connection, and the bumblebee field, we obtain the following field equations:
\begin{align}
\nonumber&R_{(\mu\nu)}(\Gamma)-\frac{1}{2}g_{\mu\nu}\bigg(R(\Gamma)+\xi B^{\alpha}B^{\beta}R_{\alpha\beta}(\Gamma)\bigg)+\\
&+2\xi\bigg(B_{(\mu}R_{\nu)\beta}(\Gamma)\bigg)B^{\beta}=\kappa^2  T_{\mu\nu};\label{Riccieq}\\
&\nabla_{\lambda}^{(\Gamma)}\bigg[\sqrt{-g}\bigg(g^{\mu\nu}+\xi B^{\mu}B^{\nu}\bigg)\bigg]=0;\label{connectioneq}\\
&\nabla_{\mu}^{(g)}B^{\mu}_{\ \nu}=-\frac{\xi}{\kappa^2}B^{\beta}R_{\beta\nu}(\Gamma)+2 V^{\prime}B_{\nu},\label{bumblebeeeq}
\end{align}
where the prime in $V'$ denotes {derivative} of $V$ with respect to its argument. Here $T_{\mu\nu}$ is given by $T_{\mu\nu}=T_{\mu\nu}^{B}+T_{\mu\nu}^{M}$, where we have defined
\begin{eqnarray}
T_{\mu\nu}^{B}&=& B_{\mu\sigma}B_{\nu}^{\ \sigma}-\frac{1}{4}g_{\mu\nu}B^{\alpha}_{\ \sigma}B^{\sigma}_{\ \alpha}-V g_{\mu\nu}+2V^{\prime}B_{\mu}B_{\nu}.
\end{eqnarray}

\subsection{Solving the connection equation}

Upon the identification
\begin{equation}\label{auxiliarymetric}
\sqrt{-h}h^{\mu\nu}=\sqrt{-g}(g^{\mu\nu}+\xi B^{\mu}B^{\nu}),
\end{equation}
one can follow \cite{BeltranJimenez:2017doy} to show that Eq.(\ref{connectioneq}) is equivalent to Eq.(\ref{conneqRBG}) up to an irrelevant projective mode. Thus, as explained above, the connection can be written as the Levi-Civita connection of $h_{\mu\nu}$ . In order to find the explicit relation between the space-time metric $g_{\mu\nu}$ and the metric $h_{\mu\nu}$,  let us rewrite (\ref{auxiliarymetric}) in matrix form as
\begin{equation}\label{auxiliarymatrix}
\sqrt{-h}\hat{h}^{-1}=\sqrt{-g}\hat{g}^{-1}\bigg(\hat{I}+\xi \hat{BB}\bigg),
\end{equation}
where the hat denotes a matrix, such that $\hat{h}^{-1}$ and $\hat{h}$ are the matrix representations of $h^{\mu\nu}$ and  $h_{\mu\nu}$, respectively. Taking the determinant of (\ref{auxiliarymatrix}) we find that $
{h}={g}\det{({I}+\xi {BB})}$ (\ref{auxiliarymetric}), and plugging this back into (\ref{auxiliarymetric}) we arrive at
\begin{equation}
h^{\mu\nu}=\frac{1}{\sqrt{\det{({I}+\xi {BB})}}}\bigg(g^{\mu\nu}+\xi B^{\mu}B^{\nu}\bigg) \ .
\label{cch}
\end{equation}
By inverting the above relation, we also have that
\begin{equation}
h_{\mu\nu}=\sqrt{\det{({I}+\xi {BB})}}\bigg(g_{\mu\nu}-\frac{\xi}{\det{({I}+\xi {BB})}}B_{\mu}B_{\nu}\bigg).
\label{ch}
\end{equation}
Using that $\det{({I}+\xi {BB})}=1+\xi X$, with $X\equiv B^\mu B_\mu$, we can finally write
\begin{eqnarray}
h^{\mu\nu}&=&\frac{1}{\sqrt{1+\xi X}}\bigg(g^{\mu\nu}+\xi B^{\mu}B^{\nu}\bigg),\label{cch1}\\
h_{\mu\nu}&=&\sqrt{1+\xi X}\bigg(g_{\mu\nu}-\frac{\xi}{1+\xi X}B_{\mu}B_{\nu}\bigg).
\label{ch1}
\end{eqnarray}
From this last result, one finds that 
\begin{equation}\label{eq:gmn_0}
g_{\mu\nu}=\frac{1}{\sqrt{1+\xi X}}h_{\mu\nu}+\frac{\xi}{1+\xi X}B_{\mu}B_{\nu} \ ,
\end{equation}
which provides an algebraic relation between $g_{\mu\nu}$ with $h_{\mu\nu}$ and $B_\mu$, though the scalar $X\equiv g^{\mu\nu}B_\mu B_\nu$ still contains an explicit dependence on $g^{\mu\nu}$. This dependence can be eliminated by noting that $Y\equiv h^{\mu\nu}B_\mu B_\nu=X\sqrt{1+\xi X}$, which completely solves the problem. Hence, we can integrate $g_{\mu\nu}$  out of the action by performing a field-redefinition  in terms of $h_{\mu\nu}$ and $B_\mu$. This is useful in order to physically interpret the different elements that contribute to the space-time metric $g_{\mu\nu}$. In fact, 
using \ref{eq:gmn_0}, we can write $g^{\mu\nu}$ in terms of $h^{\mu\nu}$ and $B_\mu$ in $\mathcal{U}_B$ and $\mathcal{S}_m$, thus finding the Einstein-frame representation of the action (\ref{bumblebee}), given by 
\begin{eqnarray}\label{EinsteinBumblebee}
\tilde{\mathcal{S}}_{BEF}&=&\int d^{4} x \sqrt{-h}\frac{1}{2 \kappa^{2}} R(h)+\overline{\mathcal{S}}_{m}\left(h_{\mu \nu}, B_{\mu}, \Psi\right),
\end{eqnarray} 
with 
\begin{align}
\nonumber&\overline{\mathcal{S}}_{m}\left(h_{\mu \nu}, B_{\mu}, \Psi_i\right)\equiv{\mathcal{S}}_{m}\big(g^{\mu\nu}(h_{\mu \nu},X,B_\mu), B_{\mu}, \Psi_i\big)-\\
\nonumber&-\frac{1}{4} \bar{B}^{\mu \nu} B_{\mu \nu}-V\left(\bar{B}^{\mu} B_{\mu} \pm b^{2}\right)+\bar{V}\left(B_{\mu}, X, h_{\mu \nu}\right)
\end{align}
where any barred tensor indicates that its indices are raised with  $h^{\mu\nu}$.

In this new (Einstein frame) representation it becomes apparent that the bumblebee model can be interpreted as GR coupled to a modified matter sector in which all the matter fields couple to the bumblebee, which also presents new self-interactions encoded in the $\bar V$ term. According to this, the metric $h_{\mu\nu}$ satisfies the Einstein equations coupled to a highly non-linear matter sector. This means that $h_{\mu\nu}$ will depart from the Minkowski metric only in regions where the Newtonian and post-Newtonian effects are expected to be relevant, {\it i.e.} regions with a strong gravitational field. As a result, as it follows from (\ref{eq:gmn_0}), the metric $g_{\mu\nu}$ will not only describe the two propagating degrees of freedom of the gravitational field through $h_{\mu\nu}$, but it will also encode information on the local value of the bumblebee field via a conformal factor and a disformal term proportional to $B_{\mu}B_{\nu} $. 

From the decomposition (\ref{eq:gmn_0}) and the fact that the independent connection satisfies $\nabla^\Gamma_\alpha h_{\mu\nu}=0$, it follows that the non-metricity tensor $Q_{\alpha\mu\nu}=\nabla^\Gamma_\alpha g_{\mu\nu}$ is {non-trivial} and entirely due to the derivatives of the bumblebee field. Since this field is expected to have a non-trivial VEV that spontaneously breaks Lorentz invariance, this is an example of a gravitationally generated non-metricity tensor that can develop a VEV. In contrast, in RBGs with minimally coupled matter, the non-metricity is associated to derivatives of the stress-energy tensor of the matter fields, which vanish in vacuum. A constant background of non-metricity was assumed in \cite{Foster:2016uui}, and experimental constraints to all its possible effective couplings to fermions and photons were derived from Lorentz violation searches in Earth laboratories. Since minimally coupled matter fields do not couple explicitly to non-metricity\footnote{Indeed as we will see later, although the new couplings that arise in the matter sector have a relation with non-metricity, rather than coupling explicitly to the non-metricity tensor of the theory, the matter fields in this model couple to the source of non-metricity instead.}, these constraints do not apply to our model. However, we note that constraints on Lorentz-violating couplings such as those in the Standard Model Extension  \cite{MISSING} could translate into constraints on the bumblebee non-minimal coupling $\xi$. Further work in this direction is currently in progress.

\subsection{Metric and bumblebee equations}

Let us now continue with the exploration of the field equations (\ref{Riccieq}) and (\ref{bumblebeeeq}). By taking the trace of (\ref{Riccieq}), a relation between the scalar curvature and the trace of the stress-energy tensor $T=g^{\mu\nu}T_{\mu\nu}$ can be found { in the form}
\begin{equation}
R(\Gamma)=-\kappa^2 T,\label{Rs}
\end{equation}
which exactly matches the relation in GR. By contracting with one and two $B^{\mu}$ fields, we can also find the relations
\begin{align}
\nonumber &B^{\mu}R_{\mu\nu}(\Gamma)=\frac{1}{1+\xi X}\bigg(\kappa^2B^{\mu}T_{\mu\nu}-\\
&-\frac{1}{2}\left(\kappa^2T+\xi R_{\alpha\beta}B^\alpha B^\beta\right)B_{\nu}\bigg),\label{Rs3}\\
&B^{\mu}B^{\nu}R_{\mu\nu}(\Gamma)=\frac{\kappa^2}{2+3\xi X}\big(-TX+2B^{\mu}B^{\nu}T_{\mu\nu}\big)\label{Rs2}
\end{align}
 respectively, where $X\equiv g^{\mu\nu}B_{\mu}B_{\nu}$. { Substituting} the second of the above equations into the first {one}, we finally obtain
\begin{eqnarray}
\nonumber B^{\alpha}R_{\alpha\nu}(\Gamma)&=&\frac{\kappa^2}{1+\xi X}\bigg(B^{\alpha}T_{\alpha\nu}-\frac{1}{2+3\xi X}\bigg[(1+\xi X)T+\\
&+&\xi T_{\alpha\beta}B^\alpha B^\beta\bigg]B_{\nu}\bigg).\label{Rs4}
\end{eqnarray}
 
The above results can be plugged back into the metric field equations (\ref{Riccieq}) to obtain an expression for $R_{\mu\nu}(h)$ which only involves $T_{\mu\nu}$, $B_\mu$ and the metric $g_{\alpha\beta}$ which, recall, can be explicitly written in terms of $h_{\mu\nu}$ and $B_\mu$. Therefore, the field equations for $h_{\mu\nu}$ can be written in Einstein like form, though their explicit form is cumbersome and does not bring any useful new insight. This confirms that we can interpret the auxiliary metric as we did when (\ref{EinsteinBumblebee}) was introduced,  as the metric that  accounts for the cumulative effects of mass and energy. 

Regarding  the bumblebee field equation, using (\ref{Rs4}) we are able to get rid of the $B^\nu R_{\mu\nu}$ term in (\ref{bumblebeeeq}), thus arriving at
\begin{eqnarray}
\nonumber\nabla_{\mu}^{(g)} B^{\mu\nu}&=&\left(2V^{\prime}+\frac{\xi T}{2+3\xi X} + \frac{\xi^2 B^{\alpha}B^{\beta}T_{\alpha\beta}}{(1+\xi X)(2+3\xi X)}\right)B^{\nu}-\\
&-&\frac{\xi}{(1+\xi X)}B_{\mu}T^{\mu\nu},
\label{Proca}
\end{eqnarray}
This equation is richer than its equivalent in the metric case, for which the $\xi$ corrections are absent. The new terms induce modifications on the effective potential that depend on the presence of other matter fields, thus implying new phenomenology. In particular, the effective mass of the field may change in regions with high energy densities, potentially leading to large effective masses and chameleon-like screening mechanisms inside and around massive objects.

\section{Weak gravitational field, post-Minkowskian limit}

Having found the solution of the connection equation and the explicit formal form of the space-time metric, we are now ready to explore the limit of negligible curvature, focusing our attention on how the sources of non-metricity modify the effective theory (post-Minkowskian limit). For this purpose, let us focus on non-gravitational experiments on Earth's surface, so that one can safely neglect all the corrections to the Minkowski metric coming from Newtonian and post-Newtonian corrections. Thus we require\footnote{Recall that (\ref{EinsteinBumblebee}) implies that $h_{\mu\nu}$ formally satisfies a set of Einstein-like equations, which yields the Minkowski metric as solution whenever one considers elementary particle interactions rather than astrophysical problems.}  $h_{\mu\nu}\approx\eta_{\mu\nu}$ . Let us also consider $\xi$ to be a small coupling and study its perturbative effects. From the above relations between the auxiliary and the space-time metrics (\ref{cch1},\ref{ch1}), we can write 
\begin{eqnarray}
\nonumber g_{\mu\nu}&=&\eta_{\mu\nu}+\xi(B_{\mu}B_{\nu}-\frac{1}{2}B^{\lambda}B_{\lambda}\eta_{\mu\nu})+\mathcal{O}(\xi^2)\\
&=&\eta_{\mu\nu}+\varepsilon_{\mu\nu},\label{efmet}
\end{eqnarray}
where $B^{\lambda}B_{\lambda}=\eta^{\mu\lambda}B_{\mu}B_{\lambda}$ and $\varepsilon_{\mu\nu}=\xi(B_{\mu}B_{\nu}-\frac{1}{2}B^{\lambda}B_{\lambda}\eta_{\mu\nu})$ $+\mathcal{O}(\xi^2)$. Here we see that even when Newtonian and post-Newtonian corrections can be neglected, the space-time metric is locally departing from its Minkowskian value due to local contributions sourced by the bumblebee field $B_\mu$. Since all fields couple to the metric, as a result, in this post-Minkow- skian approximation of our theory all the matter fields will couple to the bumblebee due to the unconventional way in which the connection mediates between the geometry and the matter. 

\subsection{Effective dynamics of scalar and fermionic fields.}

We will next proceed to study the effective dynamics of scalar and spinor matter fields in this post-Minkowskian scenario. Regarding spinors, it is important to note that they provide a nonzero contribution to the connection equation via torsional terms. Those terms have been omitted in our presentation of the field equations for simplicity. A more careful analysis, along the lines of \cite{Afonso:2017bxr}, justifies our choice because in the case of bosonic fields, torsion can be trivialized by a simple choice of projective gauge. For fermions, however, the torsion picks up contributions that cannot be gauged away. However, such terms do not modify the equation satisfied by the symmetric part of the connection, which is still of the form (\ref{connectioneq}) and admits the Christoffel symbols of $h_{\mu\nu}$ as solution. Therefore, the resulting effective metric will be the same as (\ref{efmet}). The new torsional terms will appear as new fermionic contact interactions on the right-hand side of (\ref{Riccieq}) and will be Planck-scale suppressed (see {\it e.g.} \cite{Hehl:1976kj} or \cite{Jimenez:2020dpn} for the RBG case), so that we can neglect them for our purposes.

From (\ref{efmet}) it is clear that the inverse effective metric is given by
\begin{equation}
\label{efmetinv}
g^{\mu\nu}=\eta^{\mu\nu}-\varepsilon^{\mu\nu},
\end{equation}
where the indices are raised and lowered by the usual Minkowski metric. Similarly, one has $\sqrt{-g}\simeq 1-\frac{\xi}{2} B^{\mu}B_{\mu}$.
As a result, the scalar Lagrangian within this approximation is given by
\begin{eqnarray}
\label{approx}
\mathcal{L}_{sc}&=&\frac{1}{2}\sqrt{-g}(g^{\mu\nu}\partial_{\mu}\Phi\partial_{\nu}\Phi-m^2\Phi^2)\nonumber\\
&=&
\frac{1}{2}\left(1-\frac{\xi}{2} B^{\rho}B_{\rho}\right)\left\{\big[\eta^{\mu\nu}-\epsilon^{\mu\nu}\big]\partial_{\mu}\Phi\partial_{\nu}\Phi-m^2\Phi^2\right\}.\nonumber\\
\end{eqnarray}
Using the definition of $\varepsilon^{\mu\nu}$ and integrating by parts the former equation, we arrive at
\begin{eqnarray}
\nonumber\mathcal{L}_{sc}&=&-\frac{1}{2}\Phi(\Box+m^2)\Phi-\frac{\xi}{2}(B^{\mu}\partial_{\mu}\Phi)^2+\frac{m^2}{4}\xi\Phi^2B^{\mu}B_{\mu}+\\
&+&\nonumber\mathcal{O}(\xi^2),\\
\nonumber &=&-\frac{1}{2}\Phi(\Box+m^2)\Phi+\frac{\xi}{2}\Phi\Big[B^{\mu}B^{\nu}\partial_{\mu}\partial_{\nu}+\\
\nonumber&+& \big(B^\mu (\partial_\nu B^\nu)+B^\nu (\partial_\nu B^\mu)\big)\partial_\mu+\frac{m^2}{2} B^2\Big]\Phi+\mathcal{O}(\xi^2),\\
\label{31}
\end{eqnarray}
where $\Box\equiv\eta^{\mu\nu}\partial_{\mu}\partial_{\nu}$. The interaction terms between the bumblebee and the scalar field naturally stem from special features of this metric-affine model, as explained previously. 
As the bumblebee field develops a non-trivial VEV, we see that the last two terms in (\ref{31}) carry coefficients for Lorentz violation, which will be discussed in more detail later.

Regarding spinor fields, we start with the Hermitian Lagrangian
\begin{equation}
\label{spin}{\cal L}_{sp}=\sqrt{-g}\Big[\frac{i}{2}e_a{}^\mu\left(\bar{\Psi}\gamma^a\nabla^{(\Gamma)}_{\mu}\Psi-(\nabla^{(\Gamma)}_{\mu}\bar{\Psi})\gamma^a\Psi\right)-m\bar{\Psi}\Psi
\Big],
\end{equation}
where $e_a{}^\mu$ is the tetrad field satisfying $e_a{}^\mu e_b{}^\nu g_{\mu\nu}=\eta_{ab}$. Given that the connection $\Gamma$ is just the Levi-Civita connection of $h_{\mu\nu}$, in the weak field regime $h_{\mu\nu}\approx\eta_{\mu\nu}$ we can approximate $\nabla^\Gamma_\mu \Psi\approx\partial_\mu\Psi$ up to Planck scale suppressed and $\mathcal{O}(\xi^2)$ torsion corrections (see {\it e.g.} \cite{Delhom:2019wir}). Taking this into account and using that 
\begin{equation}
e_{a}{}^\mu=\delta_a{}^\mu-\frac{1}{2} \varepsilon_{a}^\mu=\delta_{a}{}^\mu+\frac{\xi}{2}\left(\frac{1}{2} B^{2} \delta_{a}{}^\mu-B_{a} B^\mu\right),
\end{equation}
we find that the spinor Lagrangian can be written as
\begin{eqnarray}
\nonumber\mathcal{L}_{sp}&=&\bar\Psi(i \gamma^\mu \partial_\mu-m) \Psi-\\
\nonumber&-&\frac{\xi}{2} \bar\Psi\Bigg[\frac{i}{2} B^{2} \gamma^\mu \partial_{\mu}+i B^\mu B^{\nu} \gamma_{\mu} \partial_{\nu}+\frac{i}{2}\Big(B_{\alpha}\left(\partial_{\mu} B^{\alpha}\right)+\\
\nonumber&+&B^{\nu}\left(\partial_{\nu} B_{\mu}\right)+(\partial_\alpha B^\alpha) B_{\mu}\Big) \gamma^{\mu}-m B^{2}\Bigg] \Psi+\mathcal{O}(\xi^2).\\
\label{SpinorLagPert}
\end{eqnarray}
Before moving forward, it is worth calling attention to the fact that the combination of all the complex terms present in Eq. (\ref{SpinorLagPert}) result in a Hermitian Lagrangian as a consequence of Eq. (\ref{spin}).  Similarly to the scalar sector, the $\mathcal{O}(\xi)$ terms in the above Lagrangian will contribute to Lorentz violation coefficients when the bumblebee acquires a VEV. We will next proceed to analyze their physical implications.

\section{Lorentz-violating coefficients}

Let us now outline the physical effects related to a spontaneous breaking of Lorentz symmetry by the bumblebee VEV $\langle B_\mu\rangle=b_\mu$ in a weak gravitational field. Generally, observables which couple to $b_{\mu}$ will be sensitive to the spontaneous breaking of Lorentz symmetry by the bumblebee field. Since the present model displays non-minimal couplings between the bumblebee field and the matter sources through the non-Riemannian part of the connection, there will arise several Lorentz-violating (LV) coefficients in the effective matter sector once the affine connection has been integrated out.

 A straightforward and simple method to examine the effects of the Lorentz violation is to consider the VEV being a fixed-norm vector or, similarly as in Minkowski space, to take all their components to be constant. Such a choice is reasonable because we are dealing with laboratory experiments. Nonetheless, for the sake of completeness, the non-constant  $b_{\mu}$ case is presented in \ref{app1}. Therefore, from now on we restrict our analysis to the standard $b_{\mu}$ constant case \cite{KosMew}. We note that the allowed values for $b_{\mu}$ have been restricted from various experiments, see {\it e.g.}\cite{datatables} and references therein. At the same time, let us note that the LV coefficients generated by the non-metric part of the connection in metric-affine bumblebee gravity are analog to some of the parameters of the LV vector sector of the Standard Model Extension. Therefore, data from table D15 in \cite{datatable} allows us to constrain the norm of $b_{\mu}$ to be less than $10^{-43}$\, GeV.
 
\subsection{Scalar field}
Let us first study the Lagrangian for a scalar field propagating on top of the bumblebee background, which takes the form
\begin{eqnarray}
\nonumber\mathcal{L}_{sc}&=&-\frac{1}{2}\Phi(\Box+m^2)\Phi+\frac{\xi}{2}\Phi\Big[(s^{\mu\nu}\partial_{\mu}\partial_{\nu})+\frac{1}{2}m^2 b^2\Big]\Phi+\\
&+&\mathcal{O}(\xi^2),\label{ScalarLagBackground}
\end{eqnarray}
where $s^{\mu\nu}=\xi b^{\mu}b^{\nu}$. The $\mathcal{O}(\xi)$ terms will typically induce LV coefficients through the VEV of the bumblebee field. The $s^{\mu\nu}$ term constitutes a modification of the standard kinetic term which can be encoded in an effective metric for the scalar field of the form $g_{\rm eff}^{\mu\nu}=\eta^{\mu\nu}-\xi s^{\mu\nu}$. Hence, a ``wrong'' signature of the LV coefficient $s^{\mu\nu}$ could trigger ghost-like instabilities around strong enough bumblebee backgrounds. Note however that in that case, the perturbative expansion would break down since $\xi b^2$ would be $\mathcal{O}(1)$, and a full non-perturbative analysis would be required. The correction to the mass term in (\ref{31}) can also be encoded in an effective mass of the form $m_{\rm eff}^2=m^2(1-(\xi/2) b^2)$ which could also trigger tachyonic-like instabilities for a space-like bumblebee VEV (again non-perturbative effects could play a non-negligible role). 

In order to explore potential instabilities in more detail, let us analyze the particular cases of a time- and space-like constant $b_\mu$ by working out their respective dispersion relations. Starting with a time-like VEV $b_{\mu}=[b,0,0,0]$ we find the dispersion relation
\begin{equation}\label{disprelscalartimelike}
E^2=\vec{p}^2+m^2+\xi b^2 \left(\vec{p}^2+\frac{1}{2}m^2\right)+\mathcal{O}(\xi^2).
\end{equation}
This dispersion relation is healthy for positive values of $\xi$ but could potentially develop ghost- and tachyonic-like instabilities for negative values of $\xi$, although in this case the higher-order terms might be relevant to the discussion. Considering a space-like choice for $b_\mu=[0,\vec{b}]$ we end up with
\begin{equation}\label{disprelscalarspacelike}
E^2=\vec{p}^2+m^2+\xi \left(\frac{1}{2} m^2 b^2+(\vec{b}\cdot\vec{p})^2\right) +\mathcal{O}(\xi^2).
\end{equation}
Again, this dispersion relation is healthy for positive values of $\xi$. For negative values of $\xi$ a tachyonic-like instability as well as a ghost-like instability  (in directions which are non-orthogonal to $\vec{b}$) could potentially arise, and again the higher-order terms might be relevant to the discussion if they appear.

\subsection{Dirac field}

Let us now turn our attention to the spin $1/2$ fields. To explore the physics of our interest in a more convenient way, we will work with the decomposition of LV coefficients that is more commonly used in the literature \cite{Colladay:1996iz,Colladay:1998fq}. To that end, let us rewrite the weak-field spinor action \ref{SpinorLagPert} as follows
 \begin{equation}
\mathcal{L}_{sp}=\bar{\Psi}\left(i\Gamma^{\mu}\partial_{\mu}- M\right)\Psi,
\label{SpinorLagShort}
\end{equation}
where ${\Gamma}^{\mu}$ and ${M}$ are elements of the 16-dimensional Clifford algebra defined by the Dirac gamma matrices. We can thus expand them in the usual basis of this algebra as
\begin{equation}
\begin{split}
{\Gamma}^{\mu}&={e}^{\mu}I+(\delta^\mu{}_\alpha+{c}^{\mu}{}_{\alpha})\gamma^{\alpha}+{d}^{\mu}{}_{\alpha}\gamma_{5}\gamma^{\alpha}+i{f}^{\mu}\gamma_{5}+\\
&+\frac{1}{2}{g}^{\mu}{}_{\lambda\alpha}\sigma^{\lambda\alpha},\\
{M}&={m}_{\rm eff}I+{a}_{\mu}\gamma^{\mu}+{k}_{\mu}\gamma^{\mu}\gamma_{5}+\frac{1}{2}l_{\mu\nu}\sigma^{\mu\nu},
\label{LVSpinorDecomp}
\end{split}
\end{equation} 
where ${c}^{\mu}{}_{\alpha}$, ${d}^{\mu}{}_{\alpha}$, ${e}^{\mu}$, ${f}^{\mu}$, ${g}^{\mu}{}_{\lambda\alpha}$, ${a}_{\mu}$, ${k}_{\mu}$ and $l_{\mu\nu}$ are LV coefficients. Comparing \ref{SpinorLagPert} to \ref{SpinorLagShort} and \ref{LVSpinorDecomp}, we find the non-zero LV coefficients
\begin{eqnarray}
{c}^{\mu}{}_{\alpha}&=&-\frac{\xi}{2}\Big(\frac{1}{2}b^2\delta^{\mu}{}_\alpha+b^\mu b_\alpha\Big),\\
{m}_{\rm eff}&=&m\left(1-\frac{\xi}{2} b^2\right).
\label{coeff}
\end{eqnarray}
We see that within the metric-affine bumblebee model, the LV coefficients that appear provide a modification of the fermionic mass through $m_{\rm eff}$ {and} a modification of the standard kinetic term through ${c}^{\mu}_{\alpha}$. 
 In general, these will introduce modifications in the dispersion relation of spin $1/2$ fields. To that end, we first notice that the modified Dirac equation is
 \begin{equation}\label{SpinorEqShort}
\left(i\Gamma^{\mu}\partial_{\mu}- M\right)\Psi=0\ , 
\end{equation}
 and multiplying on the left by $\left(i\Gamma^{\mu}\partial_{\mu}+M\right)$ we arrive at
\begin{align}
\begin{split}
\big(-\Gamma^\mu\Gamma^\nu\partial_{\mu}\partial_{\nu}+i[M,\Gamma^\mu]\partial_{\mu}-M^2\big) \Psi=0 \ .
\end{split}
\end{align} 
By using now the relations
\begin{align}
&\{\Gamma^\mu,\Gamma^\nu\}=2\eta^{\mu\nu}-\xi(b^2\eta^{\mu\nu}+2b^\mu b^\nu)+\mathcal{O}(\xi^2),\\
&[ M,\Gamma^\mu]=\mathcal{O}(\xi^2),\\
& M^2=m^2\left(1-\xi b^2\right)+\mathcal{O}(\xi^2),
\end{align}
we find the following dispersion relation
\begin{align}
\nonumber&\bigg[E^{2}\big(1-\xi (b^{2}+b_0^2)\big)+ 2\xi b_{0} E (\vec{b} \cdot \vec{p}) -\\ 
&-\left(\vec{p}^{2}+m^{2}\right)\left(1-\xi b^{2}\right)-\xi(\vec{p} \cdot \vec{b})^{2}\Bigg] \Psi=0 \ .
\end{align}
 As we did for the scalar field, let us particularize for a constant bumblebee background of both time- and space-like types. Starting with a time-like VEV $b_{\mu}=[b,0,0,0]$ we get the dispersion relation
\begin{eqnarray}
E^{2}=(1+\xi b^2)(\vec{p}^2+m^2)+\mathcal{O}(\xi^2).
\label{Spinordispreltimelike}
\end{eqnarray} 
While for positive values of $\xi$ this dispersion relation is perfectly well-behaved, for negative values of $\xi$ there could be instabilities if $b^2\xi\geq1$, but higher order terms would be non-negligible in this case and \ref{Spinordispreltimelike} would not be trustable anymore. For a (constant) space-like vector $b_{\mu}=[0,\vec{b}]$ we obtain
\begin{eqnarray}
E^{2}=\vec{p}^2+m^2-\xi (\vec{p}\cdot\vec{b})^2+\mathcal{O}(\xi^2) \ .
\end{eqnarray}
For negative values of $\xi$ this is well behaved, though if it is positive then ghost-like instabilities could arise beyond the perturbative level ({see} \cite{Kostelecky:2000mm} for a discussion on the typical energy scales at which these instabilities become relevant).

\section{Summary and conclusions}

In this work we have formulated the bumblebee model within the metric-affine formalism.  By solving the equations of motion of the connection, we have been able to express  the metric as a function of this field and of an auxiliary metric which accounts for the standard effects of the gravitational interaction. To our knowledge, this is the first time that a solution for the connection in a curvature-based metric-affine gravity theory with spontaneously broken Lorentz symmetry has been found. The methods used to solve the connection are analog to those commonly employed in Ricci-based gravity theories (see {\it e.g.} \cite{Afonso:2017bxr,BeltranJimenez:2017doy,Bernal:2016lhq}), but its qualitative properties are rather different from those theories. In particular, while in RBGs the metric picks up local corrections that depend on the stress-energy tensor of the matter fields, here those corrections are entirely determined by the bumblebee field itself. This { occurs} due to the {absence} of higher-curvature terms in the action and to the non-minimal coupling of the bumblebee to gravity via the Ricci tensor.  

From our analysis it follows that, unlike in RBGs, where the non-metricity is given by gradients of the energy-momentum density, in our model it is the gradient of the bumblebee field that generates a non-vanishing non-metricity. The fact that the bumblebee has a VEV which breaks Lorentz symmetry allows for a background non-metricity that could fit with the proposal in \cite{Foster:2016uui}. This is the first gravitationally-induced non-metricity model with a VEV that we are aware of. 

An immediate effect of the metric dependence on the bumblebee field is that all the matter fields couple to it. Since this coupling is not a gauge one it can generate, for instance, the coupling of a vector to a neutral scalar. In the weak-field limit,  we have seen that the resulting theory looks like a bumblebee coupled to matter with non-linear interactions in Minkowski space. Therefore, this theory is naturally treated as an effective theory where the role of the energy scale is played by $\xi^{-1/2}$ (see {\it e.g.} \cite{Georgi}). Typically, for the VEV solution we have checked that the theory can present ghost and tachyonic-like instabilities depending on the range of the non-minimal parameter $\xi$ and the explicit form of the VEV. 
{ A natural continuation of this study consists in calculating quantum corrections in this effective theory.} The search for astrophysically relevant solutions beyond the weak-field limit presented here is also another research avenue currently under consideration. \\

\begin{acknowledgements}
This work was partially supported by Conselho
Nacional de Desenvolvimento Cient\'{\i}fico e Tecnol\'{o}gico (CNPq) and by the Spanish projects FIS2017-84440-C2-1-P (MINECO/FEDER, EU), i-LINK1215 (CSIC), and the projects SEJI/2017/042 and PROMETEO/2020/079 (Generalitat Valenciana). The work by A. Yu. P. has been supported by the CNPq project No. 301562/2019-9. PJP would like to thank the Brazilian agency CAPES for financial support (PDE/CAPES grand, process 88881.17175/2018-01) and Department of Physics and Astronomy, University of Pennsylvania, for the hospitality. A. D. is supported by a PhD contract of the program FPU 2015 (Spanish Ministry of Economy and Competitiveness) with reference FPU15/05406, and would like to thank the Departamento de F\'{\i}sica, Universidade Federal da Para\'{\i}ba for hospitality.
\end{acknowledgements}

\appendix
\section{Implications on taking $b_{\mu}$ non-constant}
\label{app1}

Let us start this appendix by considering the scalar Lagrangian, which takes the form
\begin{eqnarray}
\nonumber\mathcal{L}_{sc}&=&-\frac{1}{2}\Phi(\Box+m^2)\Phi+\frac{\xi}{2}\Phi\Big[(s^{\mu\nu}\partial_{\mu}\partial_{\nu})+ t^\mu \partial_\mu+\\
&+&\frac{1}{2}m^2 b^2\Big]\Phi+\mathcal{O}(\xi^2),
\end{eqnarray}
with $t^\mu=b^\mu(\partial_\nu b^\nu)+b^\nu(\partial_\nu b^\mu)$. Note  the presence of the additional coefficient $t_\mu$ in relation to the standard case (it vanishes for constant bumblebee VEVs). This coefficient introduces an imaginary term in the scalar dispersion relation. The modified scalar dispersion relations now looks
\begin{align}
\nonumber&E^2 - i\xi t_0 E -\Big[1+\xi\big(b_0^2-2b_0 (\vec{b}\cdot\vec{p})\big)\Big]\vec{p}^2-\Big[1+\xi\big(b_0^2-\\
&-2b_0 (\vec{b}\cdot\vec{p})-b^2/2\big)\Big]m^2-\xi (\vec{b}\cdot\vec{p})^2+i\xi \vec{t}\cdot\vec{p}=\mathcal{O}(\xi^2).\label{disprelscalar}
\end{align}
Note the existence of imaginary terms in the above dispersion relation, which only vanish if  one considers a frame where $t^{\mu}p_{\mu}=0$. This undesired property leads to complex eigenvalues of the Hamiltonian operator,  which  turns out to be non-Hermitian. As a consequence, we conclude that a non-constant $b_{\mu}$ produces an effective Minkowskian theory with serious problems in its dispersion relation, as displayed in the former equation. As pointed out in \cite{KosGra}, this justifies our choice of taking the Lorentz coefficient $b_{\mu}$ as a constant in order to avoid undesired effects such as a {non-Hermitian Hamiltonian and violation of energy conservation due to the fact that a non-constant $b_{\mu}$ would play the role of an external field whose presence would break space-time homogeneity}. 

For spinor fields, the dispersion relation is modified in relation to the constant $b_{\mu}$ as 
\begin{align}
\nonumber&\big(-\Gamma^\mu\Gamma^\nu\partial_{\mu}\partial_{\nu}-\Gamma^{\mu}(\partial_{\mu}\Gamma^{\nu})\partial_{\nu}-i\Gamma^{\mu}\partial_{\mu}M+i[M,\Gamma^\mu]\partial_{\mu}-\\
&-M^2\big) \Psi=0.
\end{align} 
By using now the relations
\begin{align}
&\{\Gamma^\mu,\Gamma^\nu\}=2\eta^{\mu\nu}-\xi(b^2\eta^{\mu\nu}+2b^\mu b^\nu)+\mathcal{O}(\xi^2),\\
&[ M,\Gamma^\mu]=\frac{i}{2} a_\alpha \sigma^{\alpha\mu}+\mathcal{O}(\xi^2),\\
& M^2=m^2\left(1-\xi b^2\right)+2m  a_\mu \gamma^\mu+\mathcal{O}(\xi^2),
\end{align}
where $\sigma^{\mu\nu}=\frac{i}{2}\left[\gamma^\mu, \gamma^\nu\right]$ we find the following dispersion relation
\begin{eqnarray}
\nonumber0&=&\,E^{2}\big(1-\xi (b^{2}+b_0^2)\big)+ E \Big(2\xi b_{0}(\vec{B} \cdot \vec{p})+\frac{i}{2} a_\alpha \sigma^{\alpha 0}-\\
\nonumber&-&i\gamma^{\mu}\gamma^{\beta}\partial_{\mu}c_{0\beta}\Big)-\left(\vec{p}^{2}+m^{2}\right)\left(1-\xi b^{2}\right)-\xi(\vec{p} \cdot \vec{b})^{2}-\\
\nonumber&-&\frac{i}{2}\xi m\gamma^{\mu}\partial_{\mu}b^2+i\gamma^{\mu}\gamma^{\alpha}\partial_{\mu}a_{\alpha}+i\gamma^{\mu}\gamma^{\beta}p_{i}\partial_{\mu}c_{i\beta}+\\
&+&\frac{i}{2} a_\alpha \sigma^{\alpha i} p_i-2m a_\mu \gamma^\mu.
\end{eqnarray}
Notice that the derivative terms dependent on $a_\mu$ are accompanied by gamma matrices and vanish if the bumblebee background varies slowly enough. {In whole analogy with} the scalar field situation, spinor fields suffer from the same instabilities.



\end{document}